\documentstyle[12pt]{article}
\newcommand{\be}{\begin{equation}}
\newcommand{\ee}{\end{equation}}

\begin{document}
\title{${\rm d+^3He}$ ELASTIC BACKWARD SCATTERING  AND THE
$^{3}{\rm He}$ STRUCTURE AT SHORT DISTANCES \\}
\author {
    A.P.KOBUSHKIN \\
  \it Bogolyubov Institute for Theoretical Physics \\
  \it Ukrainian Academy of Sciences, Kiev 143, Ukraine}
\date{}

\maketitle
\begin{abstract}
We study possible connection of the d+$^3$He backward elastic
scattering cross section and the mo\-men\-tum distribution of the
deuterons in $^3$He. With this aim the relativistic calculations in the
framework of proton exchange are performed.  We also consider possible
reaction mechanisms beyond the one nucleon exchange.
\end{abstract}

\mbox{}\\
PREPRINT ITP-99-5E

\newpage
\section{Introduction}
It was demonstrated \cite{Kob'86}, that  dp elastic
scattering at $\theta_{c.m.}=180^{\circ}$ and inclusive (d,p)
breakup with the proton registered at $\theta_{p}=0^{\circ}$ are
deeply connected. For example, the relativistic one-neutron-exchange
mechanism of Figure 1a describes well energy dependence of
differential cross section of the dp elastic scattering if
empirical effective momentum distribution of the proton in the
deuteron, extracted from exclusive (d,p) reaction data, is used. The
details of extraction procedure from excluisive data, as well as
comparison of the momentum distributions in the deuteron, extracted
from different reactions reactions see in
Refs.\cite{KobViz,AbleevNP,AbleevPZ,Kob'93,Perdr92}.  Still, there are
significant differences for spin-dependent observables in these two
reactions.

The mechanism, similar to the one-neutron-exchange,  was also
assumed to be a basic one for proton--lightest-nuclei backward
scattering \cite{Kopeliovich,Tecout,Lesnjak,Abdel,Zhusup,Lado93}.
In this case the colliding proton and nuclei are exchanged by a group
of $({\rm A}-1)$ nucleons, where ${\rm A}$ - is the atomic number. For
example, the simplest reaction of such kind, ${\rm p}^{3}{\rm He}$
backward elastic scattering, the two nucleon mechanism of Figure~1b
should be dominant. Indeed it was shown, that in the framework of the
dynamics at light cone and making use of the empirical momentum
distributions of the proton and the deuteron in the $^3{\rm He}$
\cite{dubna3He} one describes well the published
experimental data \cite{Berthet} for cross section of
this reaction \cite{Kob'93}. Besides the two nucleon exchange
mechanism of Figure~1b some other mechanisms were also discussed in
the literature (see \cite{Lado93} and Refs. there). For example a
$({\rm NN})_{^1S_0}+{\rm N}$-component of the $^3{\rm He}$ wave
function was shown to be of a great importance besides a
${\rm d+p}$-component.

In turn it was argued that in ${\rm d}+^{3}{\rm He}$ backward elastic
scattering the main reaction mechanism is connected with the
one-nucleon-exchange (ONE) of Figure~1c and only the
$({\rm NN})_{^1S_0}+{\rm N}$-component should contribute \cite{Tanifuji}.
The first data on differential cross section, $\frac{d\sigma}{d\Omega}$,
and tensor analyzing power, $T_{20}$, were recently measured at RIKEN
at $T_d=$140,~200 and 270~MeV \cite{Tanifuji}.

The aim of this paper is
to study how the ${\rm d}+^{3}{\rm He}$ backward elastic scattering is
connected with exclusive ${\rm A}(^{3}{\rm He},{\rm d})X$ reaction
thorough the ONE mechanism and make estimation for the differential
cross section at higher energy. Between possible
reaction mechanisms beyond the ONE we draw attention to that with
elastic deuteron-deuteron rescattering (ddS) of Figure~1d.
Due to identity of the deuterons  the dd elastic backward scattering
is the same as the forward one and the contribution of this diagram
should have the same energy dependence as that of ONE. Similar mechanism
for the pd backward scattering was already considered in
\cite{Gurvitz}.

\section{Minimal relativization of the $^3$He}
According
to \cite{Kob'93} we will use so-called "minimal relativization" scheme
of the relativis\-tic dynamics at infinite momentum frame (IMF) for the
$^3$He wave function. This is done by introducing internal momentum
between the deuteron and the proton in the $^3$He and substituting this
momentum in the nonrelativistic $^3$He wave function. For our purposes
it is enough to consider the d+p chanel of the $^3$He wave function.

Defining the IMF internal momentum, $\vec k$, we follow to
our analysis of the $^3He$ breakup published in
Ref.\cite{dubna3He,Anch-Kob}. First
one can define a fraction of the $^3$He momentum carried in
by the deuteron in the IMF in longitudial direction
\begin{equation}
\alpha=\frac{E_{\rm d}+d_3}{E_{\tau}+p_3},
\label{alpha}
\end{equation}
where $^3$He and deuteron momenta are assumed to be
$p=(E_{\tau},0,0,p_3)$ and $d=(E_{\rm d},\vec d_{\perp},d_3)$,
respectively. One can introduce invariant mass of virtual d+p pair
\begin{equation}
M^2_{\rm d+p}=
\frac{M^2+\vec d_{\perp}}{\alpha}+
\frac{m^2+\vec d_{\perp}}{1-\alpha},
\label{M_pd}
\end{equation}
where $M$ and $m$ are the deuteron and proton mass. For simplicity we
will put $M=2m$. In terms of the invariant mass $M_{\rm d+p}$ the
internal momentum is defined to be
\begin{eqnarray}
\vec k&=&(\vec k_{\perp}, k_3),\ \vec k_{\perp}=\vec d_{\perp},\\
k_3&=&\pm\sqrt{\lambda(M^2_{\rm d+p},M^2,m^2)/2M^2_{\rm
d+p}-k^2_{\perp}}.
\label{k}
\end{eqnarray}
In (\ref{k}) $\lambda(a,b,c)\equiv a^2+b^2+c^2-2ab-2ac-2bc$. The signs
"+" and "-" are chosen for $\alpha>\frac23$ and $\alpha<\frac23$,
respectively.

Similar to the deuteron (see, e.g., \cite{Karmanov'98}) one can connect
the d+p-channel wave function of the $^3$He,
$\Psi(\nu,\sigma_d,\sigma_p;\alpha,\vec k_{\perp})$, with $^3$He$\to$dp
vertex function,
\begin{equation}
\Psi(\nu,\sigma_d,\sigma_p;\alpha,\vec k_{\perp})=
\frac{\Gamma(\nu,\sigma_d,\sigma_p;\alpha,\vec k_{\perp})}
{M^2_{\rm dp}-M^2_{\tau}}
\label{Vertex}
\end{equation}
where $\nu$, $\sigma_d$, $\sigma_p$ and $L_3$ are magnetic
quantum numbers for the $^3$He, the deuteron, the proton and orbital
motion and $M_{\tau}$ is the $^3$He mass. In turn the d+p-channel
wave function of the $^3$He can be written in terms of S and D wave
components $u(k)$ and $w(k)$
\begin{eqnarray}
\Psi(\nu,\sigma_d,\sigma_p;\alpha,\vec k_{\perp})&=&
\sqrt{\frac{1}{4\pi}}u(k)
\left<1\frac12\sigma_d\sigma_p|\frac12\nu\right>+
\label{WF}\\
&+&w(k)\sum_{L_3,\mu}
\left<1\frac12\sigma_d\sigma_p|\frac32\mu\right>
\left<2\frac32 L_3 \mu|\frac12 \nu\right>Y_{2L_3}(\hat{k}).
\nonumber
\end{eqnarray}
In the framework of the dynamics in IMF the wave function (\ref{WF}) is
assumed to be normalized as
\begin{eqnarray}
\frac{1}{2}\int_{0}^{1}\frac{d\alpha}{\alpha(1-\alpha)}
\int d^2k_{\perp}
\frac{\varepsilon_p(k)\varepsilon_d(k)}{\varepsilon_p(k)+\varepsilon_d(k)}
\sum_{\nu,\sigma_d,\sigma_p}
\left|\Psi(\nu,\sigma_d,\sigma_p;\alpha,\vec k_{\perp})\right|^2
=1,
\label{Rel.Norm}
\end{eqnarray}
where
\begin{equation}
\varepsilon_p(k)=\sqrt{m^2+k^2},\ \varepsilon_d=\sqrt{M^2+k^2}.
\label{epsilon's}
\end{equation}

\section{The cross section in the relativistic ONE ap\-proxi\-ma\-tion}
According to the perturbation theory in IMF one gets the following
expression for the matrix element in the ONE approximation:
\begin{eqnarray}
T_{\nu\sigma_d\nu'\sigma_d'}&=&2S_{\rm dp}(4\pi)^2
\frac{\varepsilon_p(k)\varepsilon_d(k)}{\varepsilon_p(k)+\varepsilon_d(k)}
\times \nonumber \\
&\times&\sum_{\sigma_p}
\Gamma^{\dag}(\nu,\sigma_d',\sigma_p;\alpha,\vec k_{\perp})
\frac{1}{(1-\alpha)\left(M^2_{\rm dp}-M^2_{\tau}\right)}
\Gamma(\nu,\sigma_d,\sigma_p;\alpha,\vec k_{\perp}),
\label{Matrix_el}
\end{eqnarray}
where 2 is the combinatorial factor due to Bose statistics for
deuterons and $S_{\rm dp}$ is the spectroscopic factor in the $^3$He.
Using the connection between the wave function and the vertex function
(\ref{Vertex}) one can rewrite Eq.(\ref{Matrix_el}) in terms of the
wave function (\ref{WF})
\begin{eqnarray}
T_{\nu\sigma_d\nu'\sigma_d'}&=&2S_{\rm dp}(4\pi)^2
\frac{\varepsilon_p(k)\varepsilon_d(k)}{\varepsilon_p(k)+\varepsilon_d(k)}
\times \nonumber \\
&\times&\sum_{\sigma_p}
\Psi^{\dag}(\nu,\sigma_d',\sigma_p;\alpha,\vec k_{\perp})
\frac{\left(M^2_{\rm dp}-M^2_{\tau}\right)}{1-\alpha}
\Psi(\nu,\sigma_d,\sigma_p;\alpha,\vec k_{\perp}).
\label{Matrix_el.WF}
\end{eqnarray}
Simple calculations now give the following expression for the
differential cross section of the 180$^{\circ}$ elastic d+$^3$He
scattering:
\begin{eqnarray}
\left(\frac{d\sigma}{d\Omega}\right)_{\rm OP}=(2S_{\rm dp})^2
\frac{\overline{|T|^2}}{64s}= \frac{\pi^2}{12s}\left[
\frac{\varepsilon_p\varepsilon_d\left(M^2_{\rm dp}-M^2_{\tau}\right)}
{(\varepsilon_d+\varepsilon_p)(1-\alpha)}\right]^2 n^2_d(k),
\label{cs}
\end{eqnarray}
where $\sqrt{s}$ is the total energy of colliding particles in the
center-of-mass frame.
\section{ddS mechanism}
Now we will estimate the ddS of Figure 1d. Omitting the
real part of the dd elastic cross section at zero angle one gets that
appropriate amplitude will be pure imaginary. Because the ONE amplitude
is pure real one concludes that interference between these mechanisms
must be small:
\begin{equation}
\frac{d\sigma}{d\Omega}=
\left(\frac{d\sigma}{d\Omega}\right)_{\rm ONE}+
\left(\frac{d\sigma}{d\Omega}\right)_{dd-{\rm scat.}}.
\label{DDS-1}
\end{equation}
In turn the $dd$-scattering cross section is given by
\begin{equation}
\left(\frac{d\sigma}{d\Omega}\right)_{dd{\rm -scat.}}=
\frac{1}{(8\pi)^2s}
\left[S_{\rm dp}G_{\tau}^{d}(Q^2)M_{dd} \right]^2,
\label{DDS-2}
\end{equation}
where the formfactor $G_{\tau}^{d}(Q^2)$ is given by
\begin{eqnarray}
G_{\tau}^{d}(Q^2)&=&F_{000}\left(\frac{1}{3}Q\right) +
                  F_{022}\left(\frac{1}{3}Q\right)\nonumber \\
F_{lLL'}\left(Q\right)&=&\int_{0}^{\infty} dr j_l(Qr)
U_L(r)U_{L'}(r).
\label{DDS-3}
\end{eqnarray}
In (\ref{DDS-3}) $U_L(r)$ is radial wave function of the d+p
configuration in the  $^3$He with orbital quantum number $L$ and Q is
deuteron transferred momentum.

\section{Numerical calculations and discussion}
In our calculations we use momentum distribution of deuterons in the
$^3$He extracted in framework of impulse approximation from
$^{12}{\rm C}(^3{\rm He,d})X$ breakup data at zero angle
\cite{dubna3He}.  For spectroscopic number we use $S_{\rm dp}=1$.

To estimate contribution of the ddS we have to evaluate
amplitude of elastic dd-scattering at $\theta=0^{\circ}$ and expressed
it as
\begin{equation}
M_{\rm dd}(\theta=0^{\circ})=
2|p_{\rm dd}^{\rm c.m.}|\sqrt{s_{\rm dd}}\sigma^{\rm tot}_{\rm dd},
\label{Cal-1}
\end{equation}
where $p_{\rm dd}^{\rm c.m.}$ and $\sqrt{s_{\rm dd}}$ are deuteron
momentum and total energy of the dd-scattering in its center of mass
frame. As a total dd cross section we use
$\sigma^{\rm tot}_{\rm dd}=2\sigma^{\rm tot}_{\rm pd}\approx 150$~mb.

Results of our calculation are compared with RIKEN data at
Figure~2. One sees that the ddS mechanism is not
significant at this energy scale. In turn, as displayed at Figure~3
it becomes comparable with the ONE at $T_d\sim$1~GeV.

One sees that ONE is a main mechanism of the reaction. Nevertheless few
comments must be added:
\begin{itemize}
\item In the preset calculations we use maximal spectroscopic number
$S_{\rm dp}=1$. But three-body calculations with realistic NN-potential
give value $S_{\rm dp}<1$ (for example for Argon potential it was
obtained $S_{\rm dp}=0.69$ \cite{Schiavilla}).
\item We have estimated ddS in the framework of
nonrelativistic calcu\-la\-tions, but relativistic effect are
seem to be as important in it, as in the ONE.
\item We have used very naive assumptions about the amplitude of the
elastic dd-scattering. More realistic amplitude should be critical for
polarization obser\-va\-bles.
\item In the calculations presented here the "small components" of the
$^3He$ wave function were neglected.
\end{itemize}

Author is grateful to Professor~M.~Tanifuji and his collaborators for
fruitful discussion of their theoretical results connected with d+$^3$He
backward scattering, as well as to Dr.~T.~Uesaka for a discussion of
results of experimental situation in this field. He also thanks to
Prof.~E.A.~Strokovsky for useful comments to this work.


\begin{figure}
\mbox{}

\vspace{7cm}
\includegraphics{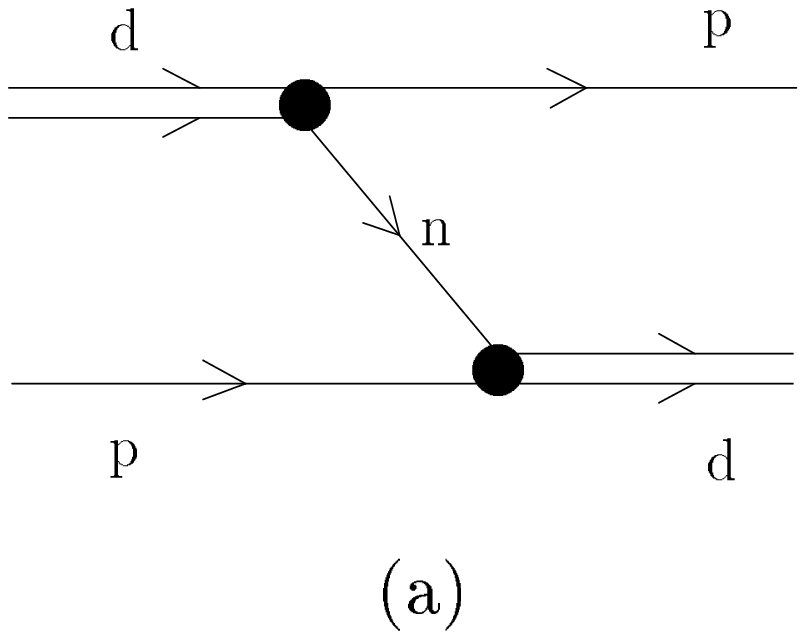}
\includegraphics{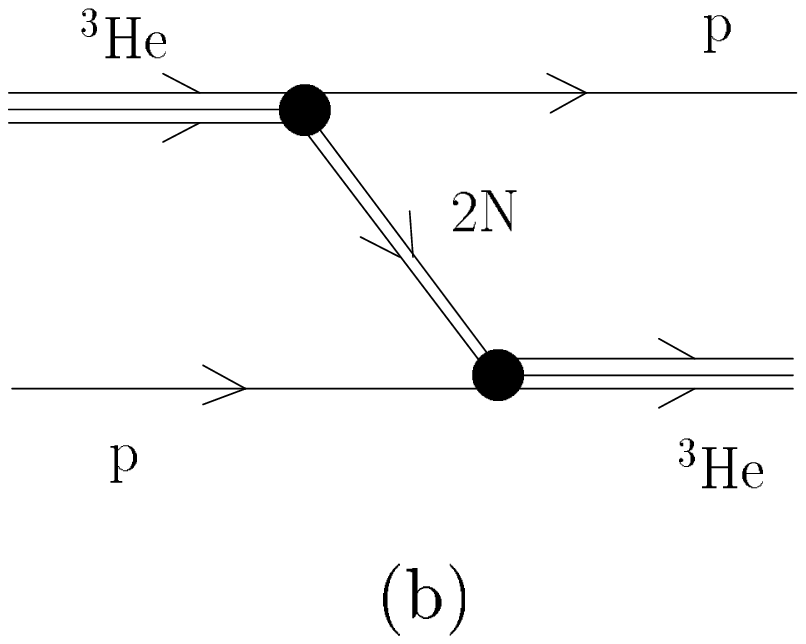}
\includegraphics{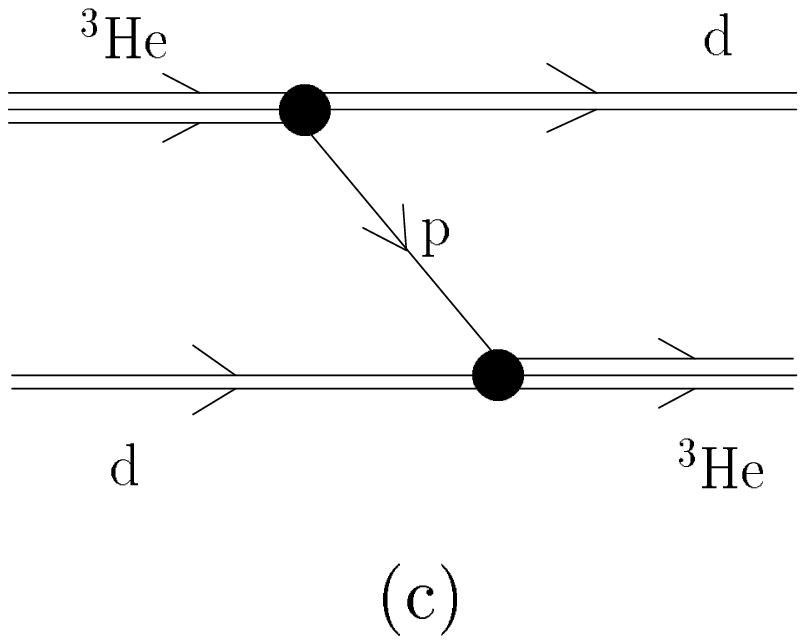}
\includegraphics{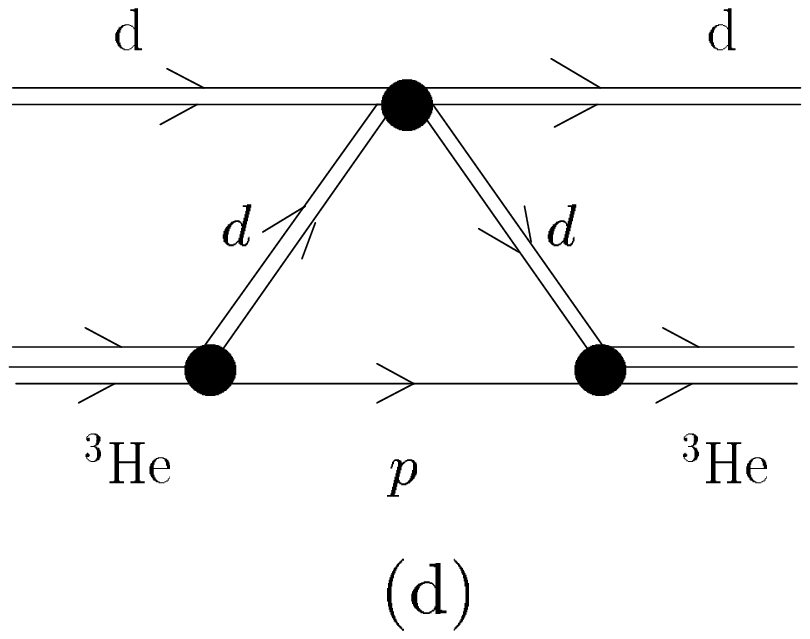}

\vspace{8cm}
\caption{One-neutron-exchange mechanism of the pd-backward
scattering (a), Two-nucleon-exchange mechanism of the p$^3$He-backward
scattering (b), One-nucleon-exchange (ONE) mechanism of the
d$^3$He-backward scattering (c) and dd-scattering (ddS) mechanism (d).}
\end{figure}
\mbox{}


\begin{figure}
\mbox{}

\vspace{8cm}
\includegraphics{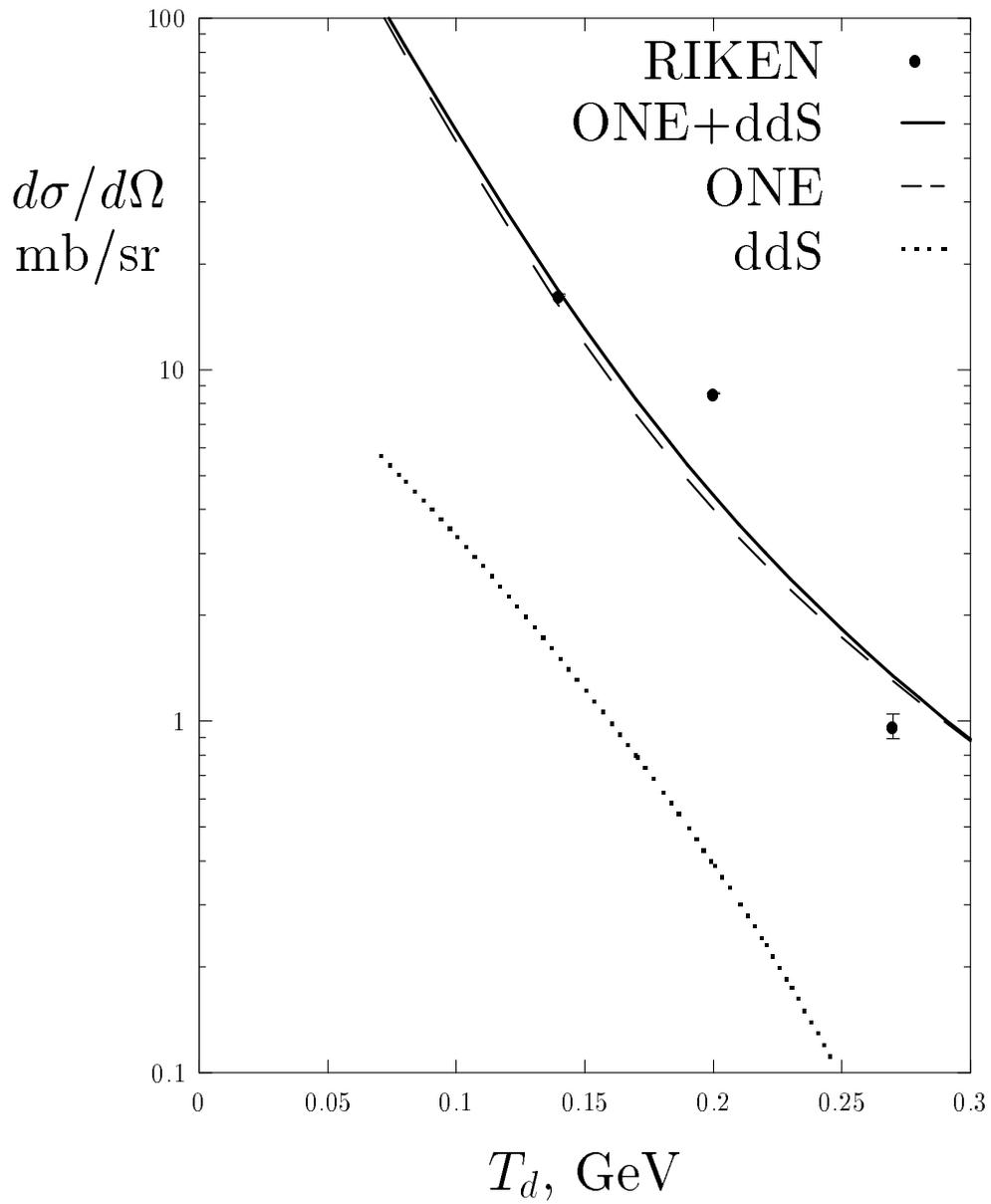}

\vspace{6cm}
\caption{Comparison of calculations for the differential
cross section with experiment [RIKEN].}
\end{figure}

\mbox{}

\begin{figure}
\mbox{}

\vspace{8cm}
\includegraphics{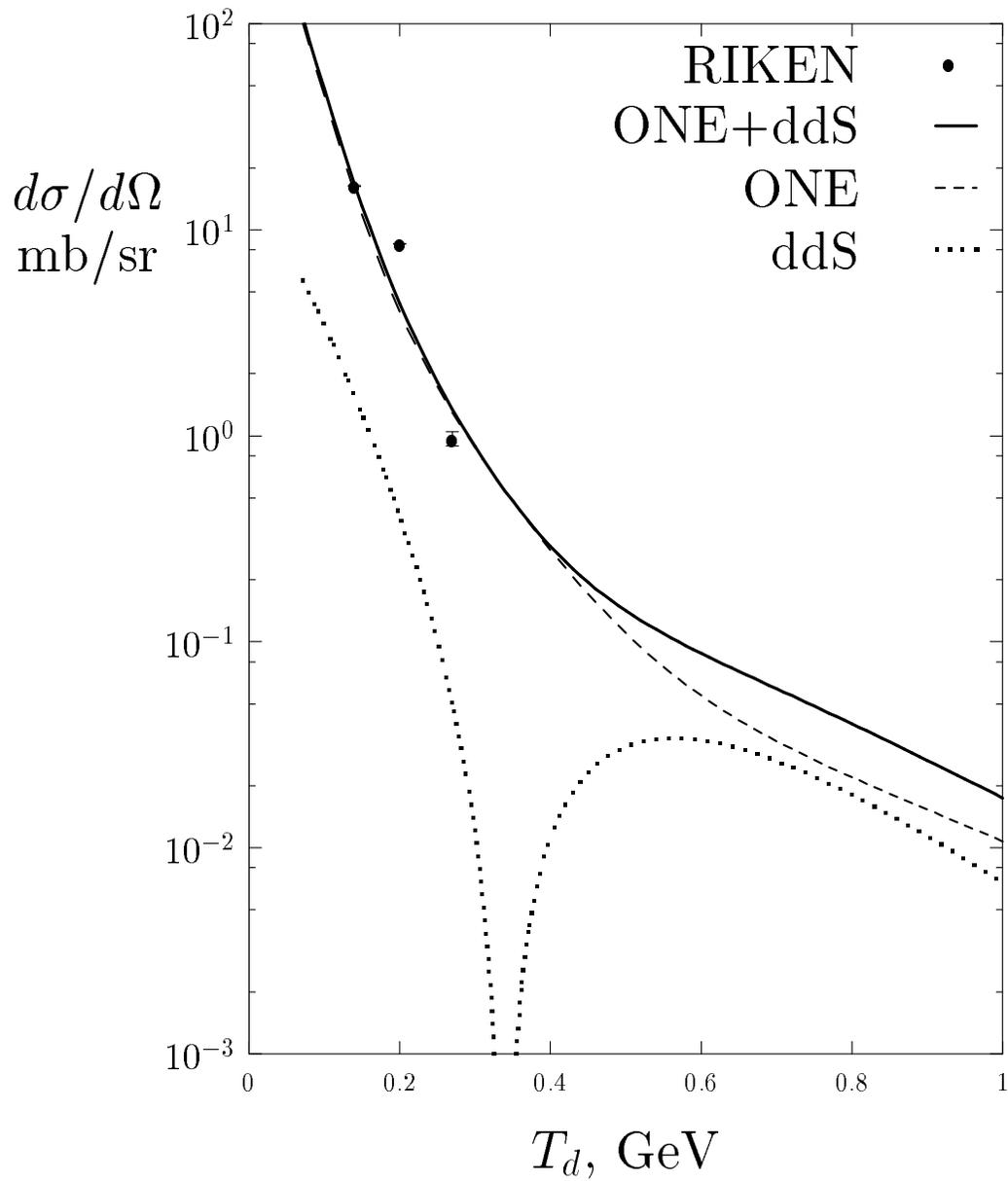}

\vspace{6cm}
\caption{Behavior the differential cross section at energy scale
of $T_d\sim$1~GeV.}
\end{figure}

\end{document}